\documentclass[twocolumn,10pt,amsmath,amssymb,apl,reprint]{revtex4-1}
\usepackage{graphicx}
\usepackage{subfigure}
\usepackage{color}

\begin{document}

\title{A self-calibrating optomechanical force sensor with femtonewton resolution} 



\author{John Melcher}
\email{john.melcher@nist.gov}	
\author{Julian Stirling}
\author{Felipe Guzm\'an Cervantes}
\author{Jon R. Pratt}
\author{Gordon A. Shaw}
\affiliation{National Institute of Standards and Technology \\ Physical Measurement Laboratory}

\date{\today}

\begin{abstract}

We report the development of an ultrasensitive optomechanical sensor designed to improve the accuracy and precision of force measurements with atomic force microscopy. The sensors reach quality factors of $4.3\times10^6$ and force resolution on the femtonewton scale at room temperature. Self-calibration of the sensor is accomplished using radiation pressure to create a reference force. Self-calibration enables {\it in situ} calibration of the sensor in extreme environments, such as cryogenic ultra-high vacuum. The senor technology presents a viable route to force measurements at the atomic scale with uncertainties below the percent level.

\end{abstract}

\maketitle

Atomic force microscopy (AFM) has proven indispensable for fundamental science at the nanoscale. The ability to measure the interaction force between a tip and sample with high precision enables measurement and manipulation at the atomic scale \cite{Ternes2008}, fundamental studies in surface chemistry \cite{DeOteyza2013}, discrimination of individual atomic species \cite{Sugimoto2007}, and interrogation of intermolecular \cite{Gross2009} and intramolecular \cite{Sweetman2011} chemical bonds. Such studies are often performed in ultra-high vacuum (UHV) at cryogenic temperatures where the limited access to the sensor prevents {\it in situ} calibration. Improving the accuracy and reliability of force measurements with AFM requires the development of {\it in-situ} calibration techniques. 

By linking the tip-sample force to the frequency of a high quality-factor ($Q$) oscillator, frequency modulation (FM-) AFM has achieved unprecedented force resolution. Oscillating a flexible microcantilever sensor in a pendulum configuration with the sample surface, Rugar {\it et al.} \cite{Rugar2004} achieved attonewton force resolution and detection of single electron spins with magnetic resonance force microscopy (MRFM). By oscillating the tip normal to the sample surface it is possible to measure the interaction force between single pairs of atoms. The challenge, however, is measuring Pauli exclusion forces between pairs of atoms in the presence of a relatively-large van der Waals background force \cite{Gross2009}. Maintaining a stable tip-sample separation at an oscillation amplitude commensurate with the decay length of the interaction requires sensors with stiffnesses on the order of $10^3$ N/m \cite{Giessibl1999, Giessibl2011, Stirling2013a}, compared to the $10^{-4}$ N/m MRFM sensor. The exquisite precision of frequency measurements from high-Q oscillators makes it possible to use a stiff sensor while preserving the force sensitivity required for measurements at the atomic scale. 

A favorite AFM sensor for UHV operation is the quartz tuning fork (QTF). QTFs are stiff sensors with mechanical quality factors at room temperature typically on the order of $10^3$ for single-tine oscillators (qPlus) \cite{NISTdisclaimer,Giessibl2004} and $10^4$ in dual-tine oscillators \cite{Qin2007, Ooe2014}. The electromechanical properties of the QTF make the sensor both self-sensing and self-actuating \cite{Giessibl2004, Qin2007}. However,limited knowledge of the sensor stiffness \cite{Simon2007, Castellanos-Gomez2009, Berger2013, Falter2014} and tip displacement \cite{Simon2007, Qin2007, Tung2010, Stirling2013} hinders the accuracy of force measurements with QTF sensors. 
 
In recent years, the coupling of optical and mechanical systems garnered considerable interest from the sensing community. The mechanical coupling results from the reversal in linear momentum of photons as they reflect from the surface of the mechanical system. Optomechanical systems allow suppression of the mechanical oscillators thermal Brownian motion \cite{Arcizet2006}. Such systems are pushing ever closer to the quantum ground state in order to enable manipulation in the quantum regime. Applications of optomechanical systems currently range from gravitational wave detectors \cite{Corbitt2007}, to optical traps \cite{Schreppler2014} and atomic clocks \cite{Wineland1984}, to AFM \cite{Kleckner2006,Hosseini2014}.

In what follows, we describe the development of an ultrasensitive optomechanical force sensor. Like the QTF, the optomechanical sensor is both self-sensing and self-actuating. In addition, the optomechanical sensor is designed to be self-calibrating, i.e., the physical mechanism needed for calibration is built in to the sensor. This is accomplished using the radiation pressure of light confined in an optical cavity to to create a reference force. The optomechanical sensors achieve quality factors on the order of $10^6$, and force resolution on the femtonewton scale at room temperature. This technology offers a viable route to atomic-scale force measurements with uncertainties below the percent level.

\begin{figure}
\centering
\includegraphics[width=0.5\textwidth]{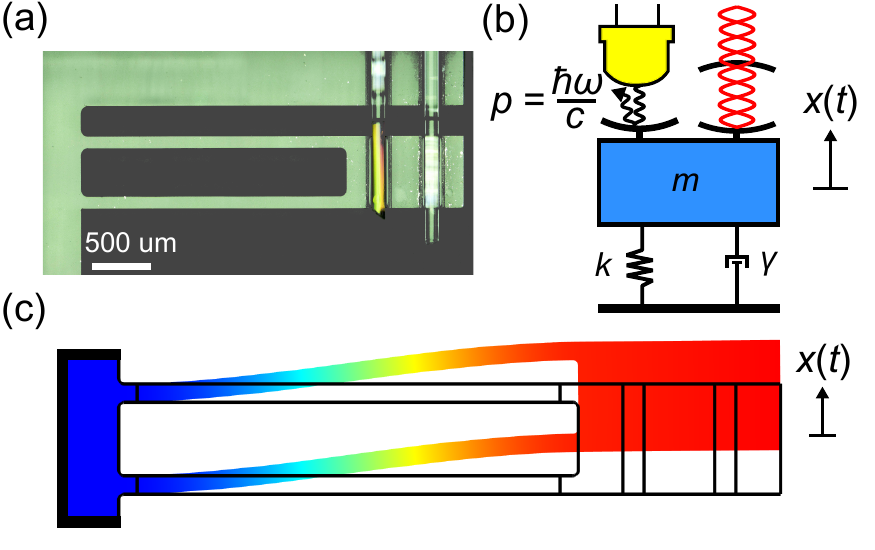}%
 \caption{\label{fig:sensor} Self-calibrating optomechanical sensor. (a) Optical image and (b) schematic of the sensor indicating a high-reflectivity mirror for actuation via radiation pressure and Fabry-Perot cavity for interferometric displacement measurement. (c) The fundamental flexure eigenmode predicted by a finite element model showing rectilinear displacement of  the proof mass. The bold line indicates the fixed boundaries.}%
\end{figure}

Tip-sample interaction forces in FM-AFM are reconstructed from the frequency shift observed while varying the tip-sample separation. Let $F$ and $z$ denote the tip-sample force and separation, respectively. The reconstructed force becomes \cite{Sader2004}:
\begin{eqnarray}\label{eq:sader}
F(z) = 2k\int_z^\infty  \left( 1 + \frac{A^{1/2}}{8 \sqrt{\pi (\zeta-z)}} \right) \Omega(\zeta)\nonumber\\ - \frac{ A^{3/2} } {\sqrt{2\left(\zeta-z \right)}}\frac{\mathrm{d} \Omega(\zeta)}{\mathrm{d}\zeta}\mathrm{d}\zeta,
\end{eqnarray}
where $\Omega = \Delta \omega/\omega_0$, $\omega_0$ and $k$ are the unperturbed frequency and stiffness of the sensor, $A$ is the oscillation amplitude. Note that in Eq. \ref{eq:sader}, it is assumed the $z$-displacement and oscillating tip-displacement are collinear. Accurately reconstruction of the tip-sample force with Eq. \ref{eq:sader} relies on calibration of the amplitude of oscillation, and, particularly, the stiffness of the sensor. 
 
Fig. \ref{fig:sensor} provides an overview of the self-calibrating optomechanical force sensor. The sensors are laser machined from a fused silica wafer. The sensor geometry is a parallelogram flexure mechanically grounded at the base with a proof mass at the distal end. There are two optical cavities between the proof mass and the support. On the left, low-coherence light from a superluminescent diode is supplied by an injection fiber. The opposing face is a high-reflectivity mirror consisting of a cleaved, gold-coated fiber that actuates sensor oscillations through radiation pressure. On the right is a Fabry-Perot optical cavity, formed by cleaved, uncoated optical fibers that is used to measure the displacement of the proof mass. The fibers are axially aligned and affixed to integrated v-grooves. The details of the displacement metrology are described by Smith et al. \cite{Smith2009}.  

The parallelogram flexure is designed to approximate the behavior of a single-degree-of-freedom (SDOF) oscillator where the transverse displacement of the proof mass in the fundamental eigenmode is approximately rectilinear (see Fig. \ref{fig:sensor}c). The design of the flexure mitigates systematic uncertainties in the force measurements by constraining the motion of the proof mass to ensure (i) accurate determination of the tip-displacement and (ii) collinearity of the tip-displacement and $z$-displacement \cite{Stirling2013}. Finally, the addition of the proof mass to the flexures causes the higher-order eigenmodes to become effectively stiff and massive \cite{Melcher2007}. Consequently, the behavior of the sensor approximates a SDOF oscillator. These characteristics allow determination of the sensor stiffness through addition of a relatively-large mass according to:
\begin{equation}\label{eq:kam}
k = \Delta m \left[ \left(\omega_0+\Delta\omega\right)^{-2}-\omega_0^{-2}\right]^{-1},
\end{equation}
where $\Delta m$ is the added mass. In this case, $\Delta m$ can easily be as large as 100 $\mathrm{\mu}$g, making it possible to calibrate with a precision microbalance. While this method is simple, accurate, and precise, it does not provide the most tractable path to self-calibration.

Calibration of the sensor can also be accomplished through the direct application of a known reference force. Wilkinson {\em et al.} \cite{Wilkinson2013} have shown that the mechanical impedance of a cantilever beam can be measured using the radiation pressure from light incident on the cantilever surface. Each individual photon carries linear momentum $p=\hbar\omega/c$, where $\hbar$ is the reduced Planck constant, $\omega$ is its frequency, and $c$ is the speed of light. The radiation force exerted by photons reflecting at normal incidence from the mirror surface at a rate $r$ becomes:
\begin{equation}\label{eq:pforce}
\frac{\mathrm{d}p}{\mathrm{d}t} = \frac{2PR}{c},
\end{equation}
where $P=\hbar \omega r$ is the power of the reference light source and $R=0.982$ is the reflectivity of the gold mirror predicted by the Fresnel equations. 

\begin{figure}
\centering
\includegraphics[width=0.5\textwidth]{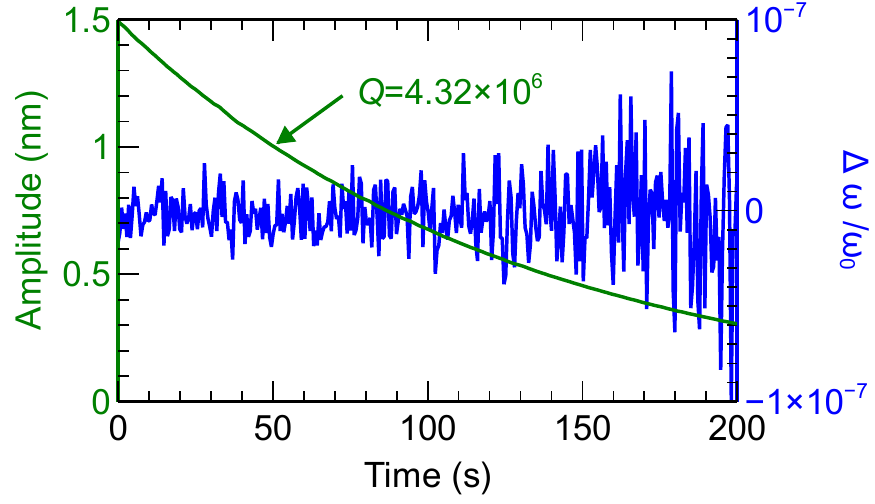}
\caption{\label{fig:ringdown} Ring-down test showing the amplitude decay and frequency stability the sensor. $Q=4.32\times10^6$ is obtained from a least-squares fit to the amplitude decay. The test was performed at room temperature in high vacuum ($10^{-4}$ Pa).}
\end{figure}

A physical mechanism for self-calibration is provided by an optical cavity where one cavity face is mechanically grounded and injects light from a reference light source. The other face is a high-reflectivity mirror aligned to the injection fiber and affixed to the proof mass. Calibration of the force sensor is then accomplished by a ring-down, ring-up sequence. The resonance frequency $\omega_0$ and quality factor $Q$ are determined from the ring-down response. Subsequently, the oscillator is self-excited by the modulating light intensity with a phase-locked loop (PLL). The sensor stiffness is determined from the steady-state amplitude of the resulting limit cycle:
\begin{equation}\label{eq:kpm}
k = \frac{2PRQ}{cA}.
\end{equation}

Fig. \ref{fig:ringdown} shows the ring-down test and frequency stability measurement for the sensor at room temperature in high vacuum ($10^{-4}$ Pa). Starting from an oscillating state, the excitation is terminated while a PLL remains locked to the oscillation. The frequency and demodulated amplitude are recorded using the PLL during the ring-down cycle. The sensor achieves a quality factor of $4.32\times10^6$, which represents a 100-fold improvement over state-of-the-art QTF sensors \cite{Ooe2014}. In addition, the relative frequency noise is on the order of $10^{-8}$ for a bandwidth of 1 Hz. The short-term frequency stability, measured by the Allan deviation, is on the order of $10^{-10}$ at a timescale of 100 s.

\begin{figure}
\centering
\includegraphics[width=0.5\textwidth]{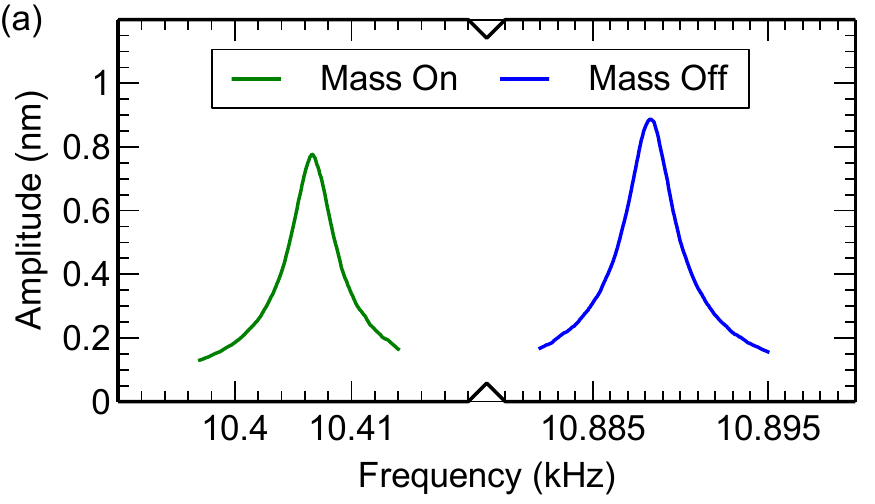}\\
\includegraphics[width=0.5\textwidth]{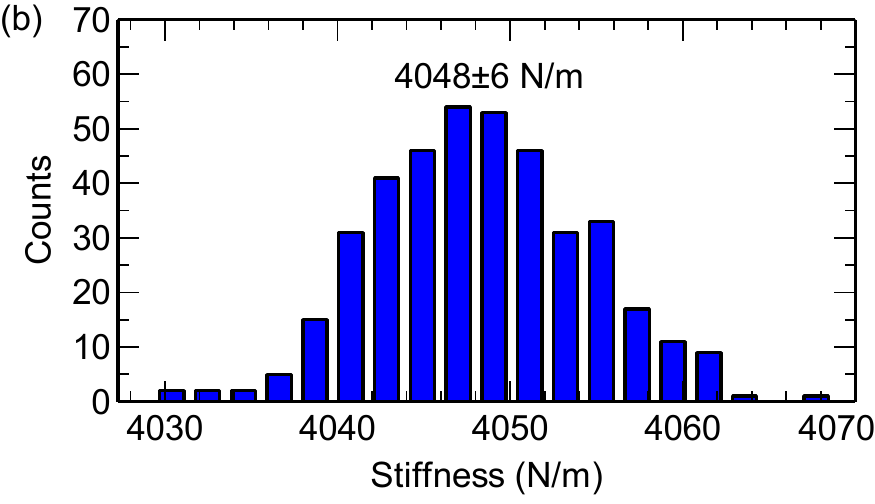}%
\caption{\label{fig:stiffness} Calibration of the force sensor stiffness. (a) Frequency response of the sensor at ambient pressure with and without an added mass  (note broken axis). (b) Stiffness calibration using the radiation pressure to create a reference force. The histogram shows repeatability of the calibration.}
\end{figure}

The stiffness of the sensor is calibrated using the added mass technique and radiation force. Prior to attaching the mirror, a Tungsten wire mass of is placed in the left v-groove. The mass of the wire $\Delta m = 82.3\pm0.6$ $\mathrm{\mu}$g is determined with a calibrated microbalance. The unperturbed frequency of the sensor $\omega_0/2\pi = 10888.27\pm0.02$ Hz and frequency shift $\Delta\omega/2\pi = -481.34\pm0.03$ Hz are determined from frequency sweeps under ambient pressure (See \ref{fig:stiffness}(a)). The stiffness is then predicted from Eq. \ref{eq:kam}. 

Next, a mirror is attached to the left v-groove allowing the sensor to be actuated with radiation pressure. The root-mean-squared (RMS) optical power of $P=2.20\pm0.02$ mW of the reference is determined with calibrated power meter \cite{Wilkinson2013}, and represents the limiting source of uncertainty. The stiffness is determined from a series of ring-down, ring-up cycles using Eq. \ref{eq:kpm} and is found to be highly repeatable (See Fig. \ref{fig:stiffness}). The results of the two calibrations are in good agreement and shown in Table \ref{tab}.
 
\begin{figure}
\centering
\includegraphics[width=0.5\textwidth]{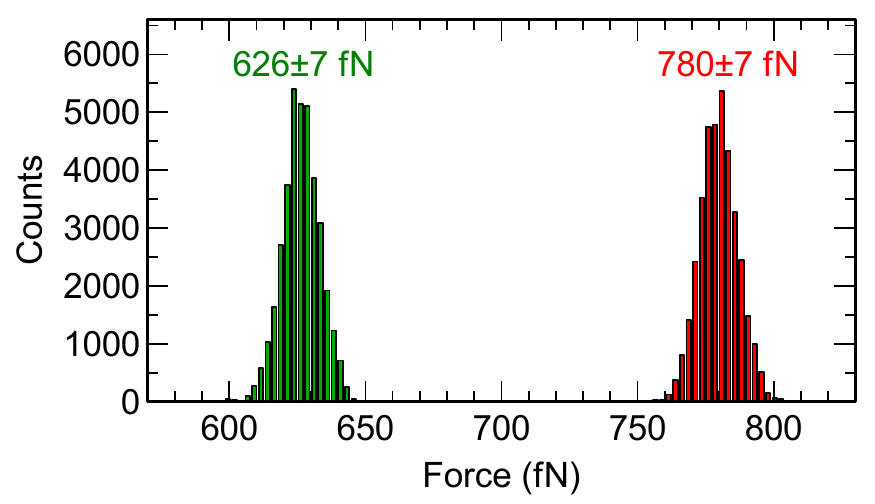}%
\caption{\label{fig:force}Measurement of the radiation force for an alternating source intensity. The standard deviation of the distributions is $\pm7$ fN and an estimated force resolution is 14 fN.}
\end{figure}

Assuming the stiffness value from the added mass calibration, we instead use the sensor measure the radiation force. For the measurements, the light source is modulated at $\omega_0$ by the PLL. The modulation amplitude is then alternated between a pair of discrete, closely-spaced RMS intensities. The radiation force is then measured by repeatedly switching the source intensity. The resulting distributions in the measured radiation force is plotted in Fig. \ref{fig:force}.

We observe two distinct distributions of $626\pm7$ fN and $780\pm7$ fN, where the uncertainty quoted represents one standard deviation of the measurements. Including the uncertainty in the calibration, the combined standard uncertainty estimate becomes $\pm9$ fN \cite{Taylor1994}. Since the distributions of the force measurements are approximately Gaussian, we conservatively estimate a force resolution of approximately 14 fN. It is important to note that this resolution is achieved with a stiff sensor that is suitable for atomic-resolution AFM, as opposed to low stiffness MRFM sensors \cite{Rugar2004} or nanowires \cite{Hosseini2014}.

\begin{table}
\caption{\label{tab} Sensor stiffness calibration results}
\begin{ruledtabular}
\begin{tabular}{ll}
\textbf{Calibration method} & \textbf{Stiffness (N/m)} \\
 Added mass & 		4066$\pm$ 31 \\
 Radiation force  & 4048$\pm$ 40 \\
\end{tabular}
\end{ruledtabular}
\end{table}

Force measurements in AFM, will benefit greatly from the development of accurate self-calibrating sensors. Perhaps the simplest example of a self-calibrating AFM sensor is cantilever probe with a built-in Fabry-Perot displacement cavity. In theory, equating the Brownian motion of the sensor to the thermal energy through the equipartition theorem \cite{Butt1995} establishes self-calibration. Force measurements at the atomic scale, however, favor the use of stiff sensors at cryogenic temperatures, which makes the thermal calibration unreliable \cite{Welker2013}. The issue of calibration at cryogenic temperatures can be overcome using radiation pressure to establish a reference force. 

In addition, the optomechanical sensor possesses an important metrological feature. From Eqs. \ref{eq:kam} and \ref{eq:kpm} it is evident that the mass, force, and laser optical power can now be linked through frequency within the SI. A reference providing any one of these quantities can then leverage the exquisite precision of frequency references \cite{Wineland1984} to form a self-calibrating system in a miniaturized package. 

We have developed an optomechanical force sensor for AFM with quality factors on the order of $10^6$ and force resolution on the femtonewton scale at room temperature. Self-calibration of the force sensor is achieved using radiation pressure to provide a reference force. The characteristics of the parallelogram flexure mitigate systematic uncertainties to improve the accuracy of force measurements in AFM. The concept of self-calibration realized through radiation pressure enables calibrations under extreme conditions such as cryogenic temperatures. Such developments in sensor technology provide a realistic path towards atomic-scale force measurements with uncertainty below the percent level.

\bibliography{ForceSensorBib.bib}

\begin{thebibliography}{32}
\bibitem [{\citenamefont {Author}(2014)}]{NISTdisclaimer}%
  \BibitemOpen
  \bibinfo {note} {Certain commercial equipment, instruments, or materials (or
  suppliers, or software, etc.) are identified in this paper to foster
  understanding. Such identification does not imply recommendation or
  endorsement by the National Institute of Standards and Technology, nor does
  it imply that the materials or equipment identified are necessarily the best
  available for the purpose.}\BibitemShut {Stop}%
\end{thebibliography}

\end{document}